# Dynamics of the freezing front during the solidification of a colloidal alumina aqueous suspension: in situ X-ray radiography, tomography and modeling

Bareggi A. [1]*, Maire E. [1], Lasalle A.[2], Deville S.[2]

[1] Université de Lyon, INSA-Lyon, MATEIS CNRS UMR5510, 7 avenue Jean Capelle, F-69621 Villeurbanne, France
[2] Laboratory of Synthesis and Functionalization of Ceramics, UMR3080 CNRS/Saint-Gobain CREE, 550 Avenue Alphonse Jauffret, BP 20224, 84306 Cavaillon Cedex, France
* Corresponding author: e-mail: bareggia@tcd.ie

ABSTRACT

Ice templating of colloidal suspension is gaining interest in material science because it offers the possibility to shape advanced materials, and in particular porous ceramics. Recent investigations on this peculiar process show that a correlation between the morphology of the frozen suspension and the velocity of the freezing front do exist. The dynamics of the freezing front of a colloidal suspension of alumina is investigated in the present study by experimental tests, finite element numerical analysis and theoretical analytical calculations. The experimental tests are carried out by in situ X-ray radiography (investigation of the dynamics of the freezing front) and tomography (investigation of the resulting morphology of the frozen suspension). The finite element model is a continuous properties model; it is used for investigating the dynamics and the shape of the freezing front. The analytical model is based on the two-phase Stefan problem. We propose a solution for the dynamics of the solidification front based on the calculation of the diffusivity as a function of the particle fraction and local temperature.

## 1. INTRODUCTION

Since 1954, ice templating, also known as freeze casting, is known as an environmental friendly and cost effective shaping process for fabricating advanced ceramic materials[1]. This shaping technique offers the possibility to introduce a certain amount of open porosity with interconnecting pore channels or pore gradients in ceramic bodies[2]. Recent publications[3,4] show the wide range of application of ice templating, including separation filters, catalyst supports, biosensors, gas distributors and implantable bone scaffolds. For introducing the pore structures in bulk ceramic bodies, suspensions with a low solid content are first frozen to obtain vehicle ice crystals which are often connected with each other in dendritic shapes, surrounded by the frozen concentrated ceramic suspension[2]. After freeze drying (removal of the ice by sublimation), porous channels remain, replicating the shape of the interconnected ice crystals. Ongoing research shows that the shape of the frozen structures and the porosity is particularly dependent on the speed of the freezing front[5].

Sintered bodies of $Al_2O_3$ with aligned channels fabricated by unidirectional ice templating of conventional aqueous ceramic slurries were reported in numerous papers[6,7,8]. These studies investigate the microstructure of the interconnected crystals of alumina, in relation to the velocity of the freezing front. The microstructure corresponding to the initial instants of solidification has been described in terms of crystal shape (R-crystals for the random structures, Z-crystals for the lamellar structures)[9]. This study also describes the structural periodicity of the lamellar texture obtained in the resulting microstructure. These investigations were carried out by high resolution in situ X-ray radiography and tomography. The purpose of the present work is to investigate the kinetics of the freezing front of colloidal suspensions of alumina by X-ray radiography and tomography. The experimental observations are compared to the results obtained with a thermal finite element model of the system (mould and colloidal suspension) and to the solution of the system of ordinary differential equation generated by the two-phase Stefan problem.

## 2. MATERIALS AND METHODS

### 2.1 Experimental setup

The experimental part of this study focuses on the use of two non-destructive imaging techniques, i.e. the high-resolution X-ray absorption radiography and tomography. We directly image the growth of the crystals in the suspension and the redistribution of the corresponding particles. X-ray attenuation radiography is a very well known technique used for years for the non-destructive imaging of samples. It is fast but the information coming from the bulk of the sample are sometimes difficult to interpret, because projected on a 2D plane. The use of tomography in materials science is newer[10]. X ray tomography combines information from a large number of X-ray radiographs taken with different viewing angles of the sample. The technique includes a computed step, i.e., a recalculation step during which a 3D map of the local absorption coefficients in each elementary volume of the sample is retrieved from the set of absorption radiographs. The reconstructed map gives an indirect image of the microstructure. X-ray tomography was performed in the present study at a voxel size of $(3.62 \text{ µm})^3$. A set of 900 projections was taken within 180°. The complete technical description of the tomograph used can be found in[11]. To summarize, this tomograph is equipped with a tungsten transmission target to produce a small size X-ray source, operated at 116 kV and 50 µA without filtering in the present study. The detector is an amorphous silicon flat panel made of 1900 * 1500 square pixels of 127 µm in lateral size.

Slurries were prepared by mixing distilled water with a small amount (0.2 wt% of the suspension) of ammonium polymethacrylate anionic dispersant (Darvan C, R. T. Vanderbilt Co., Norwalk, CT) and 32%vol of alumina powder (Ceralox SPA 0.5, Sasol, Tucson, AZ, USA, $D_{50}$=0.3 µm, SSA 8 m²/g).

Freezing experiments were performed by pouring the suspensions into a polypropylene cylindrical mould, invisible to X-rays, placed at the tip of a copper cold finger inside a cryogenic cell (see Fig. 1 -right hand side- for a scheme of the setup). The copper finger is cooled at its bottom using liquid nitrogen. This finger is 6 cm long so that the sample can be placed close to the X-ray tube for the high-resolution tomography measurement. The conduction of cold along the copper rod induces a decrease of the temperature at the bottom of the colloidal suspension, eventually leading to its freezing. The cryogenic cell is composed of a double wall PMMA envelope which prevents condensation or frost to accumulate and perturb the X-ray measurement. The atmosphere of the chamber is not refrigerated so that the suspension is forced to freeze directionally from the bottom. Freezing kinetics is controlled by a resistance heater placed at the bottom of the copper rod using the information of the temperature measured by a thermocouple at the tip of the copper rod (close to the point where the solution touches the copper). The left side of the Figure 1 shows a close up of the sample, and the arrows indicate the heat flux and the temperature at the base of the sample. These boundary conditions are used in the thermal finite element model. The experimental setup was equipped with a PID controller for setting different temperature profiles to the mould. The PID controller was connected to a PC for recording the temperature profile during the solidification process. Five different temperature profiles were used: constant, linear, and three parabolic profiles. Radiographies were acquired during freezing in order to record the evolution of the freezing front. The acquisition frequency in the radiography was 1.4 Hz, therefore the minimal time span between two images was 0.7 sec. The spatial resolution of radiography was chosen to be 11 µm. At this resolution, the entire sample was visible on the CCD screen. The maximal height of the polypropylene mould was 16 mm and the internal diameter of the mould was 3 mm. As soon as the solidification process was completed and all the information about the kinetics of the freezing front recorded in terms of temperature and radiography as a function of time, the microstructure of the frozen sample was investigated by tomography. In order to keep the full width of the sample inside the field of view of the detector, a resolution of 3.62 µm was used. Two tomographic reconstructions were necessary to analyse the morphology of the full height of each frozen sample.

## 2.2 Finite element modelling

Incremental thermo-mechanical finite element simulations were used for simulating the kinematics of the freezing front. Since the experimental configuration is fully axial symmetric, 2D axial symmetrical geometry was chosen for representing the problem. The Figure 1 (left hand side) shows a sketch of the corresponding finite element model. The equations solved in the finite element model are the conduction equation (Fourier's law) and the convection equation. A thermal conductivity was defined for each component of the system and h = 20 W/m²K was used to represent the heat transfer coefficient to air. In order to simulate the thermal behaviour of the suspension, a continuous properties model was considered. If the region of the continuous phase transition is chosen to be small in the temperature/enthalpy density diagram, the model is also a good approximation to the dynamics of substances with discontinuous phase transitions[12]. Since experimental tests show a relatively small zone of coexistence of the liquid and solid phases (and a well defined ice front), this approximation is suitable for describing the overall thermodynamic behaviour of the phase transition. In general, the thermal properties depend on local particle concentration. The results from X-ray tomography (Figure 3) show the relative fraction of the concentrated particle zone (black zones), in comparison to the pure ice (white zones). The particle concentration is limited between the nominal volumetric concentration ($\phi_{vol}$=0.32) and the maximum concentration ($\phi_p$= 0.58 is the breakthrough concentration), and between these limits the variation of the suspension properties due to the particle rich phase fraction is relatively small. According to Peppin[13], the diffusivity and the freezing temperature depend on the particle fraction, as shown in paragraph 2.3. By considering a mean value of the suspension properties as a function of the local particle concentration, it is possible to apply the proposed continuous properties finite element model. The material properties used in the modelling *i.e.* the thermal conductivity $k_T \left[\frac{W}{mK}\right]$, the specific heat capacity $Cp \left[\frac{J}{kgK}\right]$, and the density $\rho \left[\frac{kg}{m^3}\right]$, as shown by Table 1. The properties of the suspension were calculated according to the equations presented in the following section (Analytical modelling), the thermal conductivity in particular was estimated by the model of Jeffrey[14]. The main limitation of the continuous property model consists in the inability of reproducing fluctuations effects. In fact, the continuous property model does not account the rapid dendritic freezing occurring in the system; only planar and parabolic freezing front are predicted. Also, the Stefan model described in the next paragraph applies to slow planar freezing and not to the rapid dendritic freezing. Such effect can be included by particle-based simulation methods as shown by Barr[15].

## 2.3 Analytical modelling

In this paragraph, an extended Stefan model of the solidification of the colloidal suspension described in the paragraph 2.1 is presented. The problem is named after Jožef Stefan, the Slovene physicist who introduced the general class of such problems around 1890, in relation to problems of ice formation[16]. The purpose of the model is to calculate the dynamics of the freezing front depending on the temperature profiles imposed at the base of the suspension. The temperature profiles are polynomials of the sixth order and reproduce the profiles used in the experimental tests. The Stefan problem is a two-phase moving boundary problem that allows the calculation of the temperature field and the position of the interface between two phases in a melting or freezing process. The physical assumptions for the application of the Stefan problem are: conduction only, constant latent heat of fusion $L_f$, fixed melting temperature $T_m$, the interface between the two phases has no thickness, surface tension and ice nucleation are not considered (uniform freezing only).

The Stefan problem requires the estimation of the numbers of Stefan (*St*) and Lewis (*Le*). The two numbers can be estimated by using the dimensionless compressibility factor *z* as a function of particle fraction, which accounts for particle-particle interactions. The literature shows several attempts of estimating $z(\phi)$[17,18,19,20,21]. An approximation proposed by Peppin[13], valid over the entire concentration range, merges the results obtained by Woodcock[21] for concentration up to packing concentration, and the results by Carnahan and Starling[22] for low concentration. The freezing temperature is also obtained as a function of the particle fraction. The Figure 2 shows the plot of the compressibility factor z(ϕ)(upper side) and the freezing temperature as a function of the particle fraction (lower side). Details on the analytical model are provided in the Appendix. The properties of the solution, such as the heat capacity and the density, depend on the particle fraction, as shown by Figure 3 (upper side). The solution of the Stefan problem is strongly dependent on the *St* and *Le* numbers, plotted in Figure 3 (lower side) as a function of the particle fraction. The problem depends also on the estimation of the diffusivity coefficient *D,* plotted in Figure 4 as a function of the temperature of the suspension and the particle fraction. The number of Lewis indicates the influence of the Brownian diffusion on the motion of particles. At relatively small particle radius (0.5-1 μm) a Lewis number *Le = 230-250* is expected[24,25]. At this value the velocity of the freezing front is small, compared to the velocity at which the particles can migrate by Brownian diffusion, quite strong for small particles. Peppin observed that at small Lewis number, the concentration and the temperature profiles resemble those observed during alloy solidification, while for high Lewis number the weak or absent Brownian diffusion allows the formation of a porous medium ahead of the freezing front and the possibility of a morphological instability. In certain cases, the interface can become unstable owing to supercooling and particle engulfment at high front velocity[26]. If such instability is present during the solidification of a colloidal suspension, a variation in the freezing front velocity $\partial h / \partial t$ and a variation in the quantity ϕλ are induced. In other words, when the instability occurs, an acceleration (or deceleration) of the front and a discontinuity in the particle fraction occurs.

## 3. RESULTS AND DISCUSSION

The results of experimental tests consist in morphological observation and quantitative information on the dynamics of the solidification process of the colloidal suspension considered in this research. The morphology was experimentally investigated by X-ray tomography, and the results about dynamics, obtained by radiography, was compared to the results obtained by finite element modeling (FEM) and to the solution of the Stefan problem obtained by numerical solved ordinary differential equation (ODE).

## 3.1 Morphology

Four zones can be morphologically distinguished in the frozen suspension[9]. The Figure 3 shows a sample of the frozen suspension and the four zone qualitatively identified by the morphology of the crystals (on the left) and the corresponding slices obtained by X-ray tomography (black and white, on the right). The grey level images obtained by X-ray tomography were filtered with a median filter and a specifically designed threshold was used for obtaining black and white images that preserve the morphology of the crystals. In these images, the white part represents the ice, and the black part represents the alumina particle rich regions. Morphologically speaking, the crystals can be classified in "R-crystals" (random symmetry) and "Z-crystals" (vertically aligned with the axis of the mould). This distinction is important from a morphological point of view since the R-crystals reject the alumina particles in the z-direction, and the Z-crystals reject the particles in the x and y-direction. The goal of a directional ice templating technique is to obtain a structure mainly composed of Z-crystals. According to the qualitative observation[9], only the last zone of the sample (D) is purely composed of Z-crystals. The Table 2 synthetically shows the morphological zone in the frozen alumina suspensions.

In the zone between 0 and 3 mm of distance from the base of the mould, the resolution of the tomography was not good enough to image the crystals and particles distribution, therefore the actual microstructure is unknown. However, the freezing front is detectable and the measured interface velocity is particularly high (approximately 0.5-1 mm/s).

The Zone A starts as the resolution of the tomograph (voxel size: $3.62^3$ µm$^3$) allows us to obtain quantitative information on the size of the crystals. This zone is characterized by a rapid growth of the R-crystals. The velocity of the freezing front is diminishing to approximately 0.15 mm/s.

The Zone B corresponds to the transition between the zone A populated by R-crystals only and the region that shows both R-crystals and Z-crystals (zone C).

At the beginning of zone C the entrapped particles reach a local maximum corresponding to the packing of particles obtained with Z-crystals at this particular interface velocity. The recently stopped R-crystals are occupying a large fraction of the xy-plane. The thickness of this zone is strongly dependent on the temperature gradient and the cooling temperature profile. If the temperature gradient is not present (constant temperature) the zone D extends until the end of the solidification process.

The zone D represents a steady state condition, morphologically speaking, since it is composed mainly of Z-crystals that extend vertically until the end of the solidification process. Particles redistribution is occurring in the xy plane. The particle fraction is reducing almost linearly. The zone D presents thin vertical ice structures, separated by alumina packed particles, interconnected by thin horizontal alumina structures. Morphologically speaking, it is the only stable zone in the ice templating technique.

The Figure 6 shows the samples frozen using five different cooling profiles. The diagram shows the repartition of the morphological zone for each sample. It can be observed immediately that the crystals in the sample with a constant cooling profile do not reach a purely vertical morphology (pure Z-crystals are present only in zone D), and the vertical structure in this sample are quite rare. This is due to a limited resolution in the X-ray tomography. The exact position of the zone B was identified by the measurement of the particle fraction, rather than using the visual analysis. The darkening corresponds to a rapid increase of the particle rich phase fraction, followed by a morphological mutation of the crystals: the R-crystals stop growing and they are progressively being replaced by Z-crystals. The positions and the extensions of the zone B for each sample are easily identified by the curves of the particle fraction as a function of distance from the copper cold surface, shown in the second diagram of Figure 7, while the first diagram shows the temperature profiles as a function of time. In the diagram of the particle fraction the curve *par3* is not represented since it shows extremely little or no difference compared to the curve *par1*. The zone B is defined as being comprised between the local minimum and the local maximum of the particle fraction. The extension of the zone B seems to be larger in the curves than in the Figure 6. In the case considered, the solidification front is not a planar surface, but a parabolic surface that present a local maximum at the axis of the cylindrical mould. Since in Figure 6 the intersection of the conical surface with the mould is visible, the zone B appears smaller than in the tomographic reconstructions. The particle rich phase fraction at the beginning of zone A is approximately 0.57 for all the samples, and then it diminishes with different inclination to a local minimum. The behaviour of the particle fraction obtained with the constant temperature profile is quite different from the others in terms of extension of the zone B (the distance between local minimum and maximum particle fraction in the transition zone) and particle fraction at the end of the solidification process. Since the transition zone is associated with the reduction of the population of R-crystals and the growth of the population of Z-crystals, a weak transition zone, *i.e.* when the difference between the minimum and maximum particle fraction is relatively small (such as in the case of constant temperature profile) lead to an extended zone C. The sample with constant temperature profile does not show any zone with Z-crystals. In the case of constant temperature profile the velocity of the interface after the transition zone is so low that the Z-crystals can not grow. It should also be noted that the value of the constant temperature profile is close to the theoretical freezing temperature calculated in the analytical model section (Eq. 8). As the temperature profile approaches the freezing temperature, the transition zone became weaker. Deville et al.[9] observed that the transition to Z-crystals is possible only with a particular range of velocities of the freezing front. Therefore, out of this range it is not possible to describe the morphology of the frozen suspension in terms of Z-crystals and R-crystals, as done for the linear and parabolic temperature profiles used in this paper. The maximum extension of the transition zone is obtained with parabolic profiles ("par1" in particular). The steepness of parabolic profiles *par1* and *par3* maximizes the extension of the zone D by a large size of the transition zone, which reduces the extension of the zone B. A low temperature also tends to shift the transition zone backward to the cold finger, allowing more space for the steady state zone D. Not surprisingly, the condition that maximizes the zone D is a steep supercooled temperature profile, possibly parabolic.

## 3.2 Dynamics

In the experimental tests the error is generated by the resolution of the in-situ radiography (11 µm). A systematic error is also generated by the shape of the solidification front. In the initial phase of solidification, the front is a planar surface. As it strays from the copper surface, the solidification front becomes a parabolic surface. The parabolic shape of the solidification front is due to the limited insulation of the suspension from the air surrounding the polypropylene tube, almost invisible to X-ray. In order to limit the perturbation of the images in the radiography and tomography, it was chosen to provide minimal insulation. The radiography shows the solidification front in the xz plane, that becomes thicker and thicker as the distance from the copper surface increases. Finite element analysis was used for estimating the shape of the solidification front, and correcting the systematic error. The Figure 8 shows the shape of the front during three instants of the solidification with the temperature profile "par1". In order to minimize the error due to the shape of the solidification front, a reference position for the solidification front was estimated at *-2/3 err* from the upper limit, as shown in Figure 8 on the left. Due to this correction, the error on the position of the front is 11 µm. When differentiating in order to obtain the velocity of the freezing front, this error approximately represents 1% of the measured velocity.

The velocity of the solidification front was measured by in situ radiographies of the suspension during solidification. The images were recorded at a fixed time span of 0.7 seconds. The Figure 9 shows the plots of velocity of the freezing front as a function of time. The experimental results (test, solid line) are compared to the results by finite elements (FE, dashed line) and analytical model (ODE, meshed pattern line). The velocity was calculated by the position of the freezing front, considered as the projection of the front on the mould, and corrected according to the error estimation described at the beginning of the paragraph. It can be observed that the predictions by finite elements approximately follow the results by experimental tests, with a slight underestimation compared to the tests. On the other hand, the analytical model overestimates the velocity of the freezing front. A perfect fit of the analytical results with the tests was not expected. However, the finite element model based on the continuous properties model of the colloidal suspension and an analytical model based on the extended Stefan problem, that includes the calculations of the diffusion coefficient as a function of the particle fraction, provide two different approaches to the solidification problem. The parabolic profiles of temperature at the copper finger were used in order to minimize the decrease of the velocity of the freezing front as the distance between the copper finger and the front increases. This is in principle useful to obtain as soon as possible the steady state morphology of the crystals (zone D). According to the Figure 9 the velocity profiles "par1" and "par3" are the most efficient, and this is confirmed by the finite element analysis and by the analytical model. By comparing the velocities profiles of Figure 9 to the positions of zone B in the samples "par1" and "par3" in Figure 6, it can be observed that the most efficient velocity profiles shift backward the transition zone, leaving more space for developing the zone D.  This behaviour can be explained by the analytical model. For a Lewis number of 200-230 (Figure 3), Brownian diffusion is quite high. The driving force for solidification is the undercooling at the base of the suspension. At slow solidification rates the particles easily diffuse away from the interface. The temperature of the suspension ahead of the interface is always warmer than the freezing temperature. However, at faster solidification rates the concentration and concentration gradient increase at the interface. When the concentration gradient at the interface is steep enough that the gradient in the freezing temperature is larger than the temperature gradient (that in the diagram of the particle fraction in Figure 7 corresponds to a minimum of the particle fraction), the suspension ahead of the interface is below its freezing temperature (constitutionally supercooled). In analogy with binary alloys, constitutional supercooling is closely related to morphological instability, according to Davis[27].

## 4. CONCLUSIONS

This paper investigates the dynamics of the solidification front in the ice templating of alumina colloidal suspension. The methods used include experimental tests, finite element modeling and analytical modeling. The three approaches and the comparison of the results provide a deep insight in the problem of obtaining packed alumina structures in the vertical direction.

X-ray tomography allows recognizing four zones in the final product of the ice templating process. In terms of morphology, the objective of ice templating is to obtain spatially uniform lamellar structures, interconnected by bridges (interconnected Z-crystals). This condition is reached in the last zone of the frozen suspension, after a transition zone. The analytical model shows that in the transition zone, if the Lewis number is relatively high (*Le = 200-230*), the rejected particle form a packed layer, followed by a zone characterized by a variation of the morphology: the random shape crystals (R-crystals) stop growing and the Z-crystals start to develop. The extension of the Z-crystals zone depends on the size and the extension of the transition zone. By 3D reconstruction of data by X-ray tomography it is possible to plot the particle fraction as a function of the vertical position. The transition zone is identified by a two local singular points (minimum and maximum) of the particle fraction, and the distance between these singular points indicates the extension of the transition zone. By controlling the temperature at the base of the suspension it is possible to control the extension of the Z-crystals zone. A steep parabolic profile of the temperature maximizes the extension of the Z-crystals zone by a strong reduction of the transition zone (that also reduces the extension of the zone B). The lower parabolic temperature profile at the base of the suspension is associated with a shift of the transition zone towards the base of the suspension, leaving more space for the Z-crystals to develop. In terms of modeling, the thermal properties of a colloidal suspension in a solidification process depend on the particle fraction. However, results from finite elements show that a model based on constant properties is sufficient to provide reliable results. In principle, the thermal properties depend on the particle concentration, however constant properties can be used since the actual concentration range of the suspension considered in this paper is limited (0.32-0.58). A model for calculating each property should be chosen carefully in the literature, with particular attention to the specific heat capacity. The parameter that should be checked for choosing an appropriate model is the Lewis number, which represents the importance of the Brownian motion. The equations used in this paper for predicting the thermal properties are fitted on small particles (0.5-1 μm).

The analytical model provides results in terms of velocity of the freezing front. The model can be used to compare the responses of the system to different temperature profiles, thus it has to be properly tuned in order to obtain the actual velocity profile. Therefore, the analytical model can be used to provide guidelines in the choice of the temperature profile, in order to maximize the extension of the zone characterized by vertical lamellar structures (Z- crystal zone).

In conclusion, the morphology of ceramics shaped with the technique of ice templating can be partially controlled by choosing the profile of the temperature at the base of the suspension. The dynamics of the freezing process of alumina colloidal suspension can be modeled by analytical calculations and finite elements. The two approaches complementary contribute to the modeling of the problem: the analytical model estimate the thermal properties of the colloidal suspension that have to be used in the finite element model. Also, the analytical model provides the estimation of the number of Lewis *Le* and the diffusion coefficient *D*, that resulted to be the most important parameters to predict the morphological behavior of colloidal suspension[18,27]. The analytical works presented in this paper allow estimating *Le* and *D* as a function of the medium properties, the particle radius, and the concentration of alumina. Future works should investigate the impact of the dispersant on the number of Lewis.

According to the experimental tests, to the finite element analysis and to the solution of the Stefan problem, a successful approach consists in applying a cooling rate approximately between 2 and 4°C/min; as the transition zone occurs, it is worth increasing the cooling rate following a parabolic profile. Since the position of the transition zone is *a priori* unknown, it is convenient to apply a parabolic profile since the beginning of the process.

## **ACKNOWLEDGEMENTS**

Financial support was provided by the National Research Agency (ANR), project NACRE in the non-thematic BLANC programme, reference BLAN07-2_192446.

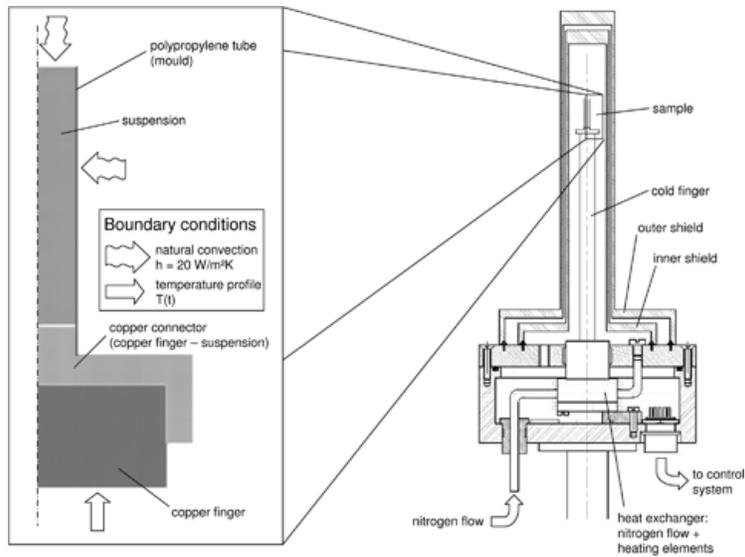

Figure 1 – On the right: sketch of the freezing device. A flux of cold nitrogen is pumped at constant rate on a ring in contact with a copper finger, in contact with the colloidal suspension. A heating element controlled by a PID controller is used for regulating the temperature at the base of the suspension. A double shield provides thermal isolation from the ambient. On the left: the geometry (cylindrical symmetry) and boundary conditions of the finite element model. The natural convection was applied to the upper boundary of the suspension and on the polypropylene mould.

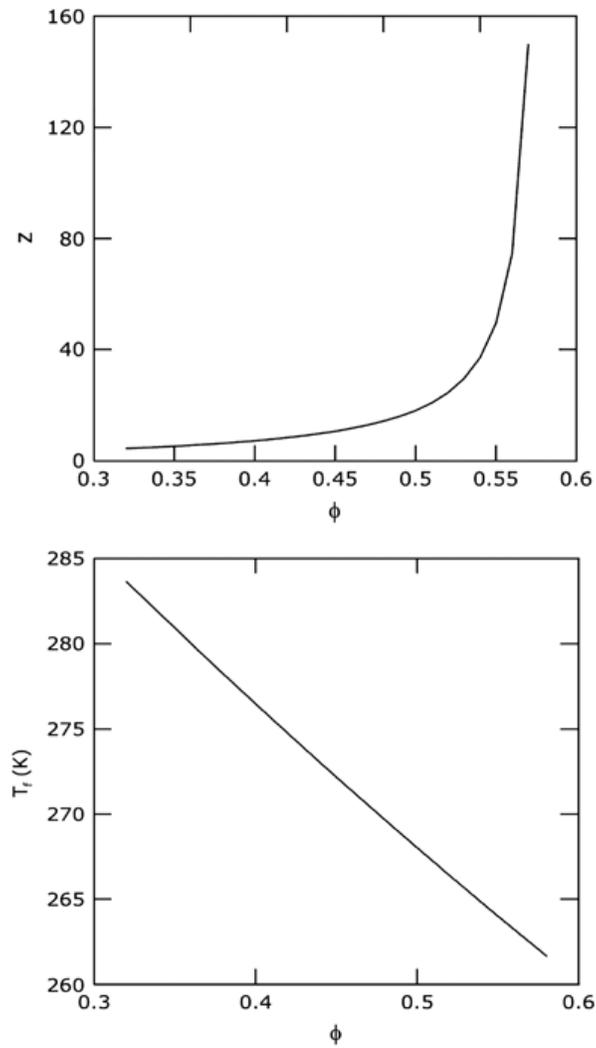

Figure 2 - The dimensionless compressibility factor $z$ (upper side), the freezing temperature $T_f$ expressed in K (lower side) as a function on the particle fraction $\Phi$.

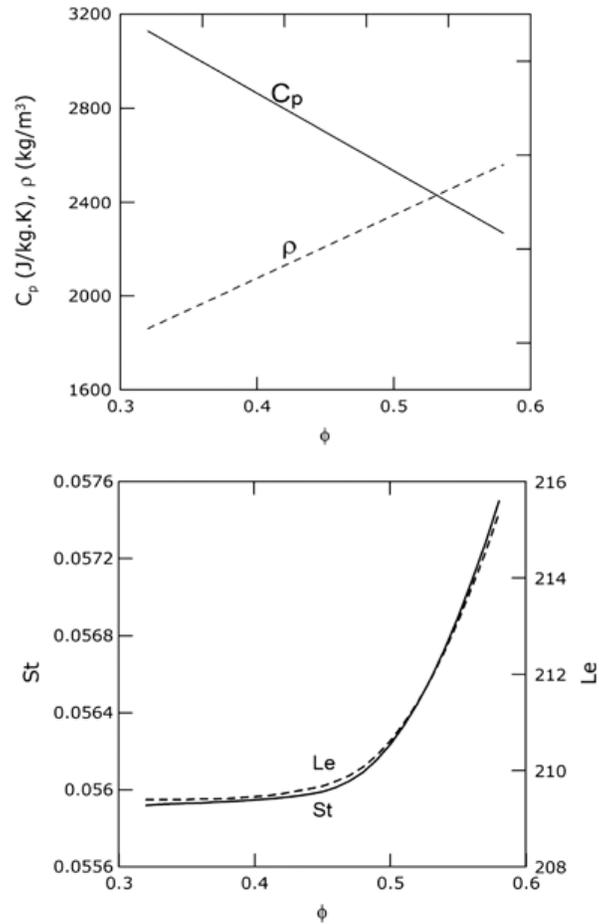

Figure 3 – The specific heat capacity $Cp$ and the density $\rho$ for the liquid phase of the suspension (upper side). The Stefan and Lewis used in the Stefan problem (lower side) as a function on the particle fraction $\Phi$.

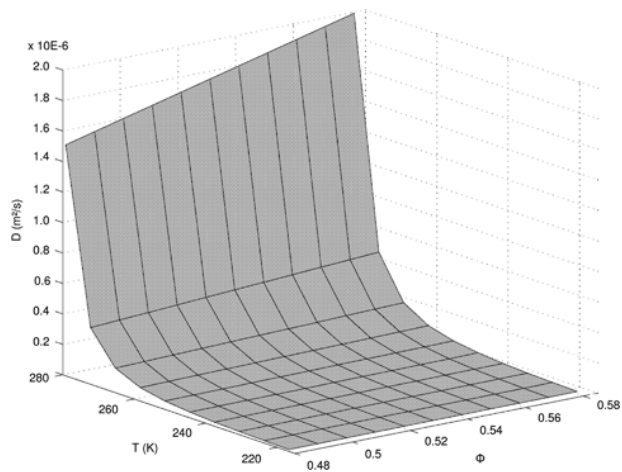

Figure 4 - The diffusivity coefficient $D$ as a function of local temperature and particle fraction.

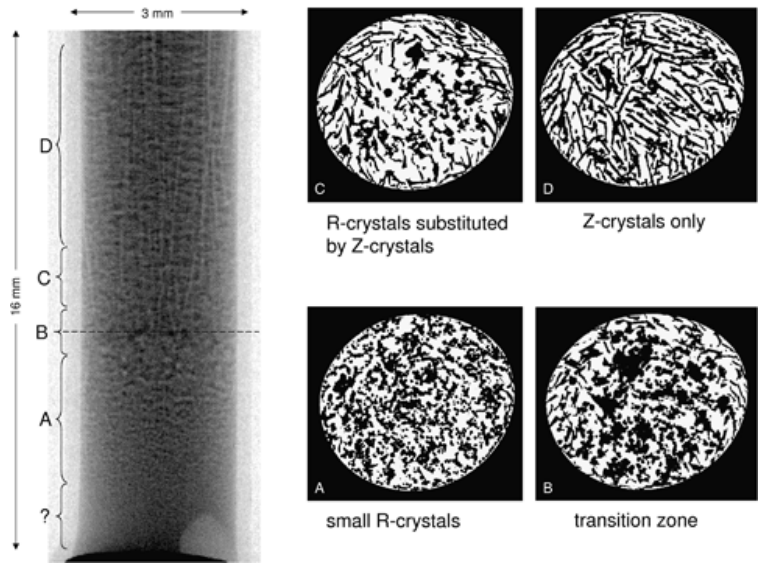

Figure 5 – The result of the radiography of the alumina sample frozen with the temperature profile "par1". The sample measures 16 mm in height and 3 mm in mean diameter. The morphological zones are indicated in the diagram on the left. The three slices on the right indicate the results of X-ray tomography. The bottom slice shows the suspension at the upper limit of zone A (small R-crystals). The middle slice shows the suspension in the zone B: R-crystals rapidly grow and few Z-crystals nucleate close to the polypropylene mould; since the zone C is approaching, the biggest R-crystals tent to group. The upper slice shows the suspension in the zone D (steady state zone), characterized by Z-crystals.

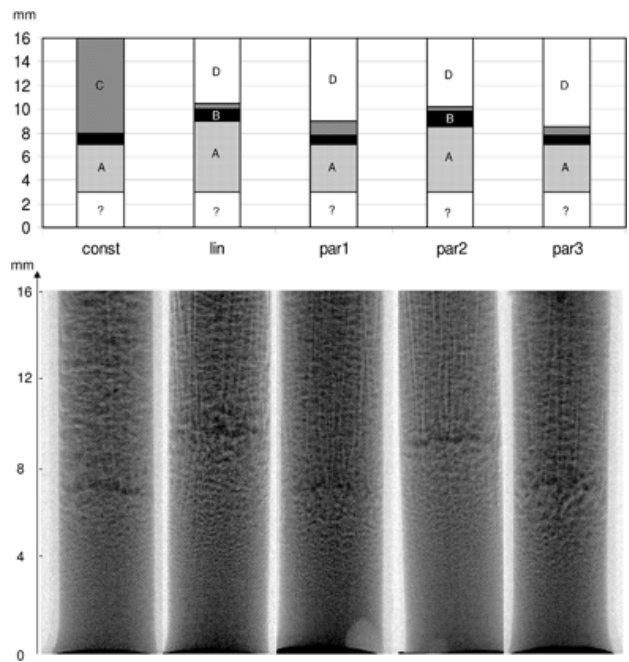

Figure 6 – The four morphological zone for the all the temperature profiles. The upper diagram shows the subdivision of each sample in the morphological zones. The lower diagram shows the results of the radiography at the end of the solidification process.

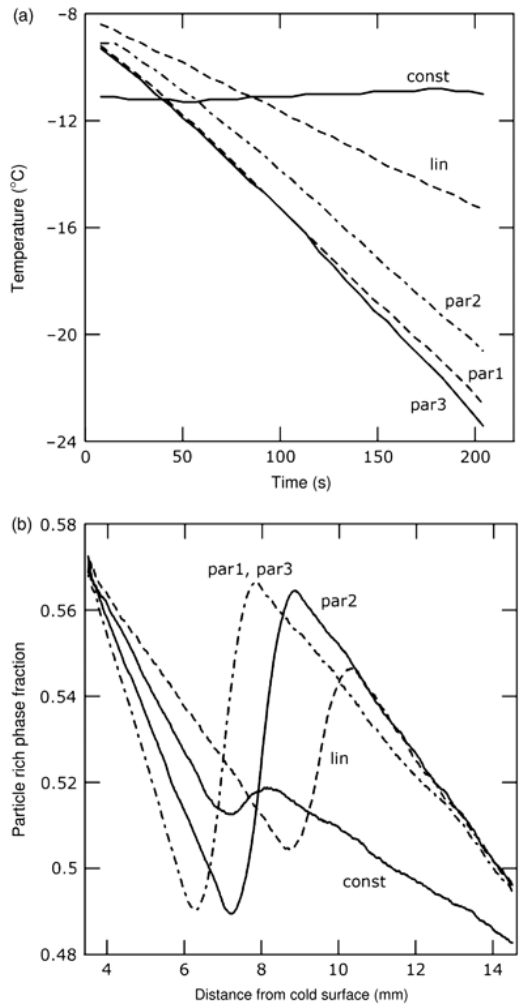

Figure 7 – The upper diagram shows the temperature profiles applied at the base of the colloidal suspension (a). The lower diagram shows the particle fraction as a function of the distance from the base of the suspension, plotted for all the temperature profiles (b).

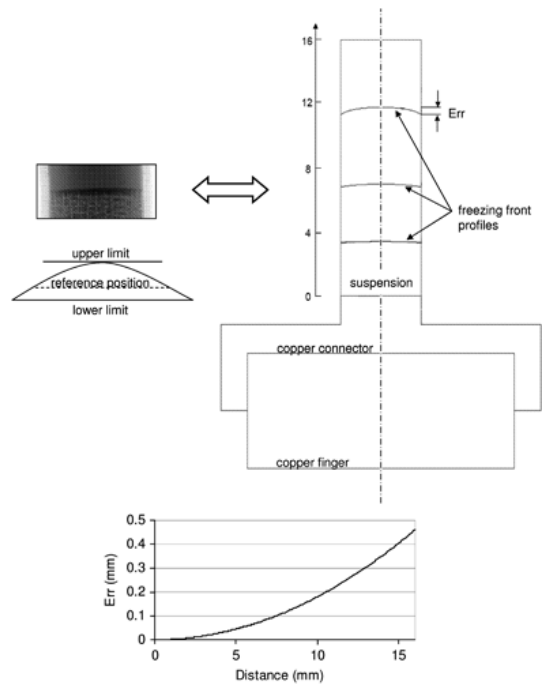

Figure 8 – The correction of the error due to the shape of the solidification front. The diagram on the right shows the prediction by finite elements of the shape of the solidification front in three instants of the process. The uncertainty generated by the shape of the freezing front presents an upper and a lower limit. The error, defined as the difference between the upper and the lower limit, is plotted in the bottom diagram.

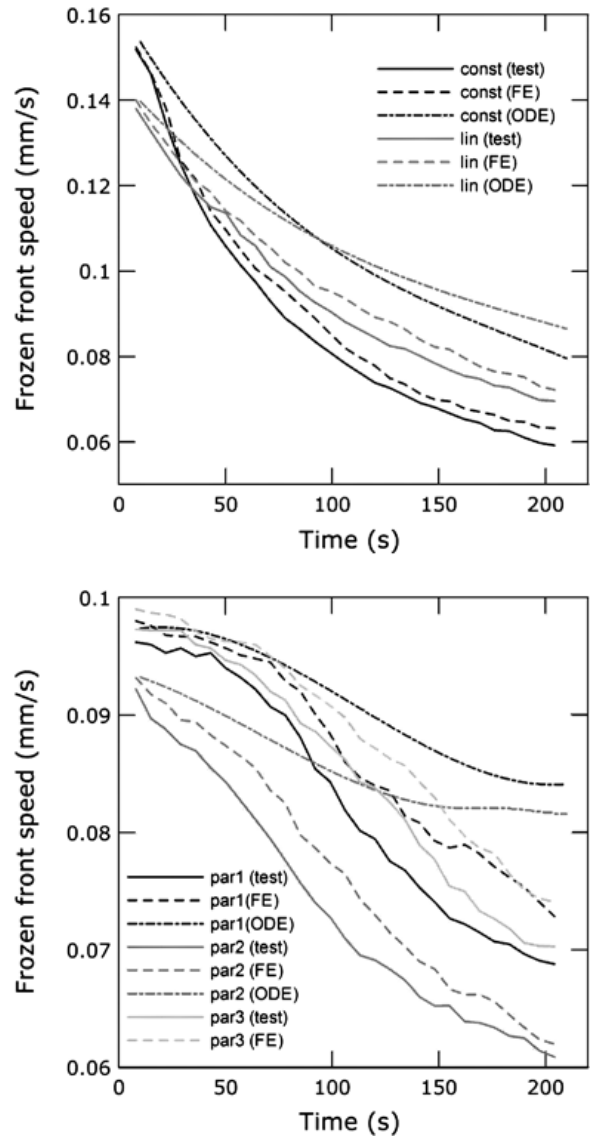

Figure 9 – The comparative diagrams of the velocity of the freezing front. The experimental results (test, solid line) are compared to the results by finite elements (FE, dashed line) and analytical model (ODE, meshed pattern line). The upper diagram shows the velocities of the freezing front when a constant or linear temperature profile is applied at the base of the suspension (a). The lower diagram shows the velocities in the cases of the parabolic temperature profiles (b). The velocity was calculated by differentiating the position of the freezing front, considered as the projection of the front on the mould, and corrected according to the error estimation described by Figure 6.

TABLES

| **Nomenclature** | |
|---|---|
| $C_p$ | Specific heat capacity of the suspension |
| $D$ | Diffusivity coefficient |
| $D_0$ | Einstein-Stokes diffusivity |
| $h(t)$ | Solidification front position |
| $Le$ | Lewis number |
| $L_f$ | Latent heat of fusion of medium (water) |
| $k_B$ | Constant of Boltzmann |
| $k_T$ | Thermal conductivity of the suspension |
| R | Radius of the alumina particle |
| $St$ | Stefan number |
| $T$ | Local temperature |
| $T_b$ | Temperature at the base of the mould |
| $T_f(\phi)$ | Freezing temperature |
| $T_m$ | Melting temperature of the medium (water) |
| $v_p$ | Volume of the alumina particle |
| $z(\phi)$ | Dimensionless compressibility factor |
| $\alpha$ | Thermal diffusivity of the suspension |
| $\eta$ | Similarity variable of the Stefan problem |
| $\vartheta$ | Dimensionless temperature |
| $\theta_b$ | Dimensionless temp. (base of the mould) |
| $\theta_f$ | Freezing dimensionless temperature |
| $\theta_m$ | Melting dimensionless temp. (water) |
| $\lambda(t)$ | Similarity variable at the interface |
| $\mu$ | Viscosity of the medium (water) |
| $\rho$ | Density of the suspension |
| $\Pi(\Phi)$ | Osmotic pressure of the suspension |
| $\phi$ | Particle fraction |
| $\phi_p$ | Packing particle fraction |
| $\phi_{vol}$ | Nominal volumetric particle fraction |

**Table 1 - Thermal properties (FEM and analytical modeling)**

|              | L-phase | S-phase | copper | polypropylene |
|--------------|---------|---------|--------|---------------|
| kT [W/mK]    | 0.6     | 0.9187  | 403    | 0.12          |
| Cp [J/Kg K]  | 2434    | 1457    | 385    | 1925          |
| ρ [Kg/m³]    | 2426    | 2218    | 8930   | 950           |

**Table 2 – Morphological zones**

| Zone | R-crystals  | Z-crystals | morphological stability |
|------|-------------|------------|-------------------------|
| A    | growing     | growing    | no                      |
| B    | stop growth | growing    | no                      |
| C    | no          | growing    | no                      |
| D    | no          | yes        | yes                     |

APPENDIX – Details on the analytical model

The Stefan problem requires the estimation of the numbers of Stefan (*St*) and Lewis (*Le*). The two numbers are defined as following:

$$St = \frac{L_f}{\rho C_p T}, \quad Le = \frac{\alpha}{D_0} \quad (Eq.1)$$

were $L_f$ is the latent heat of fusion of water, $\rho$ and $Cp$ are respectively the density and the specific heat capacity of the suspension, $\alpha$ is the thermal diffusivity of the suspension $\left(\alpha \equiv \frac{k_T}{\rho C p}\right)$, $T$ is the absolute local temperature of the suspension, and $D_0$ is the Stokes-Einstein diffusivity, defined as follows:

$$D_0 = \frac{k_B T}{6\pi R \mu} \quad (Eq.2)$$

with $k_B$ the Boltzmann's constant, $R$ the radius of the alumina particle, $\mu$ the dynamic viscosity of the carrying fluid. In order to calculate the numbers of Stefan and Lewis, the thermal properties of the suspension were estimated for the liquid and the solid phases.

The thermal conductivity was estimated by the equation of Jeffrey[14] used for modeling the thermal conductivity of the suspension (Eq.3):

$$k_{TS} = k_{fluid}\left(1 + 3C_{vol}\beta + 3\Phi_{vol}^2 \beta^2 \gamma\right)$$

Eq.3

$$\gamma = 1 + \frac{\beta}{4} + \frac{3\beta}{16}\frac{\alpha+2}{2\alpha+3}, \quad \alpha = \frac{k_{ice}}{k_{fluid}}, \quad \beta = \frac{\alpha-1}{\alpha+2}$$

With $k_{ice,\ fluid}$ the thermal conductivities of the medium (water) in the solid and in the liquid phase, and $\phi_{vol}$ the nominal volumetric concentration of the alumina particles: $k_{ice} = 2.2 \frac{W}{mK}, k_{fluid} = 0.6 \frac{W}{mK}, \Phi_{vol} = 0.32$,

The thermal conductivity of the suspension after freezing can be obtained by Eq.3 and is equal to $k_{TS} = 0.9187 \frac{W}{mK}$. The thermal conductivity of the suspension before freezing can be approximated by the thermal conductivity of the medium (liquid water)[12]: $k_{TL} = 0.6 \frac{W}{mK}$. For concentrations $C_{vol} < 0.5$, the specific heat capacity and the density of the suspension ($Cp$ and $\rho$) can be estimated by the following relation:

$$Cp, \rho_{L,S} = (1-\Phi_P)Cp, \rho_{water(L,S)} + (\Phi_P)Cp, \rho_{part} \quad \text{Eq.4}$$

Where $\phi_p=0.58$ is the packing fraction of particle in the suspension in the concentrated region between the ice crystals at ambient temperature and pressure. The Stokes-Einstein diffusivity has also been used by Peppin et al.[13] for writing the diffusion coefficient as a function of the local absolute temperature $T$ and the local particle fraction $\phi$ (Eq. 5)

$$D(\Phi,T) = D_0 \hat{D}(\Phi),$$
$$\hat{D}(\Phi) = (1-\Phi)^6 \frac{d(\Phi z)}{d\Phi} \quad \text{(Eq. 5)}$$

where $z$ is the dimensionless compressibility factor which accounts for the effect of particle-particle interactions on the osmotic pressure $\Pi(\Phi)$, which can be written as

$$\Pi(\Phi) = \frac{\Phi}{v_p} k_B T z(\Phi) \quad \text{(Eq. 6)}$$

The expression of the osmotic pressure as a function of the particle fraction is a fundamental result from statistical mechanics[29] which states that the osmotic pressure of a suspension of particles and the pressure of a perfect gas are the same functions of volume fraction if the particle-particle interaction potential is the same. Peppin[13] propose the following estimation of the compressibility factor:

$$z(\Phi) = \frac{1 + a_1\Phi + a_2\Phi^2 + a_3\Phi^3 + a_4\Phi^4}{1 - \Phi/\Phi_P}, \quad \text{(Eq. 7)}$$
$$a_1 = 4 - 1/\Phi_P, a_2 = 10 - 4/\Phi_P, a_3 = 18 - 10/\Phi_P, a_2 = 1.5/\Phi_P^5 - 18/\Phi_P$$

Peppin[13] obtain the freezing temperature by utilising the conditions for local thermodynamic equilibrium between the liquid phase and the solid phase, assuming a planar interface and isotropic stress between the two phases, and the latent heat of fusion approximately constant in the temperature range considered. From these assumptions Peppin obtains the value of the freezing temperature $T_f(\Phi)$:

$$T_f = T_m(1 + mz\phi)^{-1}$$
$$m = \frac{k_B T_m}{v_p \rho_L L_f} \quad \text{(Eq. 8)}$$

with $T_m=273K$, the melting temperature of water.

Once the thermal properties of the suspension ($k_T$, $C_p$ and $\rho$) the diffusion coefficient $D$ and the freezing temperature $T_f$ have been estimated, the Stefan problem for unidirectional solidification of the suspension of hard-sphere colloids can be formulated. The Figure 2 shows the quantities involved in the formulation of the Stefan problem as a function of ϕ. The Stefan problem is a system of two ordinary differential equations (ODE) and its solution represents the temperature field and the position $h(t)$ of the solidification front. The Equation 9 shows the ODEs of the Stefan problem with the boundary conditions.

$$\frac{\partial \Phi}{\partial t} = \frac{\partial}{\partial x} D \frac{\partial \Phi}{\partial x}$$

B.C. $\Phi = \Phi_p (x = 0, t = 0)$

$$\Phi \frac{\partial h}{\partial t} = -D \frac{\partial \Phi}{\partial x} \quad (x = t) \qquad \text{(Eq. 9, Stefan problem)}$$

and

$$\frac{\partial T}{\partial t} = \alpha \frac{\partial^2 T}{\partial x^2}$$

B.C. $T = T_b(t)(x = 0)$

$$\rho L_f \frac{\partial h}{\partial t} = k_T \left( \frac{\partial T}{\partial x} \bigg|_{x=h^-} - \frac{\partial T}{\partial x} \bigg|_{x=h^+} \right) \quad (x = h,\ T = T_f(\Phi)) \text{ Stefan condition}$$

The Stefan problem admits a similarity solution[16] with the variable

$\eta = \dfrac{x}{\sqrt{4 D_0 t}}$ , the interface position $h(t) = 2\lambda \sqrt{D_0 t}$ , and the dimensionless temperature

$$\theta = \frac{T - T_\infty}{T_f - T_\infty} \qquad \text{(Eq. 10)}$$

The formulation of the Stefan problem is

$$\frac{\partial \Phi}{\partial \eta} = -\frac{1}{2\eta} \frac{\partial}{\partial \eta} D(\Phi) \frac{\partial \Phi}{\partial \eta}$$

$$\frac{\partial \theta}{\partial \eta} = -\frac{Le}{2\eta} \frac{\partial^2 \theta}{\partial \eta^2}$$

B.C. (Eq. 11)

$\Phi = \Phi_p \quad (\eta = 0)$

$\theta = \theta_h(\eta)$ , $\Phi = \Phi_h(\eta)$ , $\Phi \lambda = \dfrac{D(\Phi)}{2} \dfrac{\partial \Phi}{\partial \eta} \bigg|_{\eta=\lambda} \quad (\eta = \lambda)$

with $\theta_h = \theta_m + \Gamma \dfrac{\Phi_h z}{1 + m\Phi_h z} \bigg|_h$, $\Gamma = \dfrac{mT_m}{T_\infty - T_f(\Phi_\infty)} \bigg|_{\Phi_\infty}$

The problem was solved by using the MATLAB ODE suite, according to the methods illustrated by Shampine and Reichelt[23]. By substituting the results for $\Phi$ in the Stephan condition $(\eta = \lambda)$, it is possible to obtain the expression of $\lambda$ for each boundary condition:

$$\lambda(t) = -\frac{Le}{2St}\left(\frac{\partial \theta}{\partial \eta}\bigg|_{\eta=\lambda-} - \frac{\partial \theta}{\partial \eta}\bigg|_{\eta=\lambda+}\right) \quad (\eta = \lambda) \quad \text{Eq. 12}$$